\RequirePackage{fix-cm}
\documentclass[smallextended]{svjour3}       
\smartqed  
\usepackage{graphicx}
\usepackage{algorithm}
\usepackage{algorithmic}
\usepackage{amssymb}
\usepackage{times}
\usepackage{epsfig}
\usepackage{graphicx}
\usepackage{amsmath}
\usepackage{amssymb}
\usepackage{caption}
\usepackage{subcaption}
\usepackage{amssymb}
\usepackage{pifont}
\usepackage{url}
\usepackage{pifont}
\usepackage{amssymb}
\usepackage[numbers]{natbib}
\usepackage[pagebackref=false,breaklinks=true,letterpaper=true,colorlinks,
bookmarks=false]{hyperref}
\usepackage{hyperref}

\usepackage{lineno}
\usepackage{multirow}
\usepackage{float}
\begin{document}

\title{Movie Recommendation System using Sentiment Analysis from Microblogging Data
}


\author{Sudhanshu Kumar       \and
        Shirsendu S. Halder \and Kanjar De \and Partha Pratim Roy 
}

\institute{Sudhanshu Kumar, Shirsendu S. Halder, Kanjar De, Partha Pratim Roy 
\at
              Department of Computer Science and Engineering \\ 
              Indian Institute of Technology, Roorkee, India. \\
              \email{sud16.dcs2016@iitr.ac.in}           
}

\date{Received: date / Accepted: date}

\maketitle

\begin{abstract}
Recommendation systems are important intelligent systems that play a vital role in providing selective information to users. Traditional approaches in recommendation systems include collaborative filtering and content-based filtering. However, these approaches have certain limitations like the necessity of prior user history and habits for performing the task of recommendation. In order to reduce the effect of such dependencies, this paper proposes a hybrid recommendation system which combines the collaborative filtering, content-based filtering with sentiment analysis of movie tweets. The movie tweets have been collected from microblogging websites to understand the current trends and user response of the movie. Experiments conducted on public database produce promising results.
\keywords{Recommendation System \and Twitter \and Collaborative filtering  \and Content-based 
filtering \and Sentiment Analysis}
\end{abstract}

\section{Introduction}
\label{intro}
In today's world, internet has become an important part of the human life. Users often face the problem of excessive available information. Recommendation systems (RS) are deployed to help users cope with this information explosion. RS are mostly used in e-commerce applications and knowledge management systems such as tourism, entertainment and online shopping portals. In this paper, we focus on RS for movies are an important source of recreation and entertainment in our life. Movie suggestions for users depend on web-based portals. Movies can be easily differentiated through their genres like comedy, thriller, animation, and action. Another possible way to categorize movies can be achieved on the basis of metadata such as year, language, director or by cast. 
Most online video-streaming services provide a number of similar movies to the user to utilising the user's previously viewed or rated history. Movie Recommendation Systems \cite{katarya2017effective,li2016two,wan2011utilizing,adomavicius2005toward, zhou2017predicting} help us to search our preferred movies and also reduce the trouble of spending a lot of time searching for favorable movies. The primary requirement of a movie recommendation system is that, it should be very reliable and provide the user with the recommendation of movies which are similar to their preferences. In recent times, with exponential increase in amount of online-data, RS are very beneficial for taking decisions in different activities of day-to-day life. RS are broadly classified into two categories: Collaborative filtering (CF) and Content-based filtering (CBF). 

It is a human tendency to take decisions on the basis of facts, predefined rules and known information which is available over the internet and this inclination of human behaviour gave rise to the concept of CF. Resnick et al. \cite{resnick1994grouplens} introduced the concept of CF in netnews, to help people find articles they liked in a huge stream of available articles. CF help people make choices based on the perspective of other people. Two users are considered like-minded when their rating for items are similar whereas in CBF \cite{van2000using}, items are suggested through the similarity among the contentual information of the items. With the advent of numerous social media platforms like Quora, Facebook, and Twitter, people are able to share their daily state of mind on the internet. Twitter \cite{abel2011analyzing, abel2013twitter, fu2014study} is one of the most popular social media platform founded in 2006 where users can express their thoughts in limited characters. The Unique Selling Proposition of Twitter is that the existing users not only receive information according to their social links, but also gain access to other user-generated information. The source of information on Twitter are called tweets which are of limited-character that keep users updated about their favorite topics, people and movies. In this paper, we propose a movie recommendation framework  by fusing hybrid and sentiment scores from MovieTweetings database.


The main contributions of the paper are as follows:
\begin{enumerate}
\item We propose a hybrid recommendation system by combining collaborative filtering and content-based filtering. 
\item Sentiment analysis is used to boost up this recommendation system.
\item A detailed analysis of proposed recommendation system is presented through extensive experiment. Finally, a qualitative as well as quantitative comparison with other baselines models is also demonstrated.

\end{enumerate}  

The rest of the paper is organized as follows: Section~\ref{sec:relwrk} summarized the related work. The proposed methodology is 
presented in Section~\ref{sec:method}. Results obtained using the proposed 
framework are shown in Section~\ref{sec:results}. Finally, the conclusion is drawn in Section~\ref{sec:conc}.

\section{Related work}
\label{sec:relwrk}

Recommender Systems are the most effective knowledge
management systems that help users to filter unusable data
and deliver personalized ideas based on their past historicial data and similar items which user are looking over the internet. Many RS have been developed over the
past decades. These systems used different approaches like
CF, CBF, hybrid and  
sentiment analysis system to recommend items. These are discussed 
as follows.
%

\subsection{Collaborative (CF), Content-based (CBF) and Hybrid filtering}
Various RS approaches have been proposed in literature for recommending items \cite{sembium2017recommending}. Primordial use of collaborative filtering was introduced in \cite{goldberg1992using} which proposed a search system based on document contents and responses collected from other users. A RS for e-commerce applications by combining the pre-purchase and post-purchase ratings was suggested by Guo et al. \cite{guo2013new}. Soleymani et al. \cite{soleymani2008affective} worked on an affective 
ranking of movie scenes, based on user's emotion and video content 
based features. They found that the movie scenes were correlated with the 
user’s self-assessed arousal and valence. The peripheral physiological 
signals were utilized to categorize and rank video contents. 

Yang et al. \cite{yang2009cares} inferred 
implicit ratings from the number of pages the users read. As many pages users 
read, the more the users are assumed to like the documents. This concept is helpful to overcome the cold start problem in CF.     
Optimizing a RS is an ill-posed problem. Researchers have proposed several optimization algorithms such as gray wolf optimization \cite{katarya2016recommender}, artificial bee colony \cite{hsu2012personalized}, 
particle swarm optimization \cite{ujjin2003particle} and genetic algorithms \cite{bobadilla2011improving}, etc. Katarya et al. 
\cite{katarya2016recommender} developed a movie recommendation system based on 
collaborative filtering which utilizes the bio-inspired gray wolf optimizer 
and fuzzy c-mean (FCM) clustering techniques. The Gray wolf 
optimizer was applied to obtain initial position of the cluster. The movie 
ratings were further predicted based on user's historical data and similarity of users. They improved the existing framework in \cite{katarya2018movie} proposing ABC-KM (artificial bee colony and 
k-mean cluster) framework for collaborative movie recommendation system to 
reduce the scalability and cold start complication. This hybrid cluster and 
optimization technique combination showed better accuracy in movie prediction compared to existing frameworks. 

Content-based filtering \cite{wasid2015particle,lops2011content,philip2014application, viard2018movie} is one of the 
most widely used and researched recommendation system paradigm. This RS approach is based on the description 
of the item and a profile of the user’s preferences. In \cite{nascimento2011source}, the authors discussed about the discriminative power of words. They deduce that the title and abstract are multiple times stronger than the body text and use a weighting scheme of the title, abstract and the body text to retrieve relevant documents. Cantador et al. \cite{cantador2010content} make use of user and item profiles, described in terms of weighted lists of social tags to provide music recommendations. Meteran et al. \cite{van2000using} proposed a Personalized Recommender System (PRES) to suggest articles for home improvement where the similarity between the user profile vector and a document was determined by using the combination of TF-IDF and the cosine similarity. Goossen et al. \cite{goossen2011news} proposed a new method for recommending news items
based on TF-IDF and a domain ontology i.e. CF-IDF. The performance of this method outperformed the TF-IDF approach on several measures such as accuracy, recall, and the $F_1$-measures when tested, evaluated and implemented on Athena framework.  

Recent research has demonstrated that hybrid approach \cite{porcel2012hybrid,romadhony2013online, spiegel2009hybrid, aslanian2016hybrid, burke2002hybrid} is more effective than traditional approaches. The main advantage of hybrid systems are that they combine multiple recommendation techniques to mitigate the drawbacks of individual techniques. In \cite{melville2002content}, the authors have developed a content-boosted CF System which used pure content-based features in a collaborative framework. This system further improved the prediction, first-rater and the sparsity problem.
Zhang et al. \cite{zhang2015hybrid} developed a framework based on user    
recommender interaction that takes input from the user, recommends N items to the 
user, and records user choice until none of the recommended items favors. In \cite{noguera2012mobile}, the authors developed a mobile recommender system that combines a hybrid recommendation engine and a mobile 3D GIS architecture. For testing the proposed framework, twenty seven users were selected having an age range of 24-48 years. To evaluate the performance of the system users were instructed to find restaurant, bars and accommodation while walking and driving along a motorway. The user feedbacks demonstrated competent performance by the 3D map-based interface that also overcomes the limited display size of most mobile devices.

\subsection{Sentiment analysis}
Sentiment analysis \cite{prabowo2009sentiment,cambria2016affective,medhat2014sentiment,
ravi2015survey,gauba2017prediction, yadava2017analysis} is a widely used technique by researchers to acquire people opinions. In \cite{gallege2016towards}, sentiment 
analysis has been used for rating products for online software services. Their research enhances both CBF and CF algorithms, 
using external reviews such as sentiment analysis and subjective logic. 
Sentiment analysis technique has been used in 
\cite{loria2014textblob} to calculate the polarity and confidence of review 
sentences. The authors in \cite{hutto2014vader} proposed Valence Aware Dictionary and Sentiment Reasoner (VADER) model for sentiment analysis. Lexical
features were combined for five general rules that embody grammatical and syntactical conventions for expressing and emphasizing sentiment intensity. An
$F_1$ score of 0.96 was recorded for classifying the tweets into positive, neutral, and negative classes. In \cite{anto2016product}, the author proposed automatic feedback 
technique on the basis of data collected from Twitter. Different classifiers such as Support Vector Machine, Naive Bayes, and Maximum Entropy were used on twitter comments. 
The authors in \cite{rosa2015music} proposed music recommendation framework for mobile 
devices where recommendations of songs for a user were based on the mood of user's 
sentiment intensity. The studies were performed on 200 participants (100 men and 
100 women) to fill out their musical preference choice in his or her profile. 
Later, the participant's profile was analyzed and the results showed 91\% 
user satisfaction rating. In \cite{li2016intelligent}, the author proposed 
KBridge framework to solve the cold start problem in CF 
system. Sentiment analysis was also used for microblogging posts in this 
framework and polarity score of the post was assigned in 1 to 5 rating. The 
result showed an enhanced recommendation system by bridging the gap between user
communication knowledge between social networking sites and program watching over TV.
The author in \cite{leung2006integrating} proposed a rating inference approach to transform textual reviews into ratings to enable easy integration of sentiment analysis and CF.

Our proposed model is a hybrid recommender system whose results are boosted using sentiment analysis score. Experimental evaluations, both quantitative and qualitative demonstrate the validity and effectiveness of our method.

\section{Proposed system} 
\label{sec:method}

\begin{figure}[!h!t!b]
	\begin{center}
		\includegraphics[width=.9\linewidth]{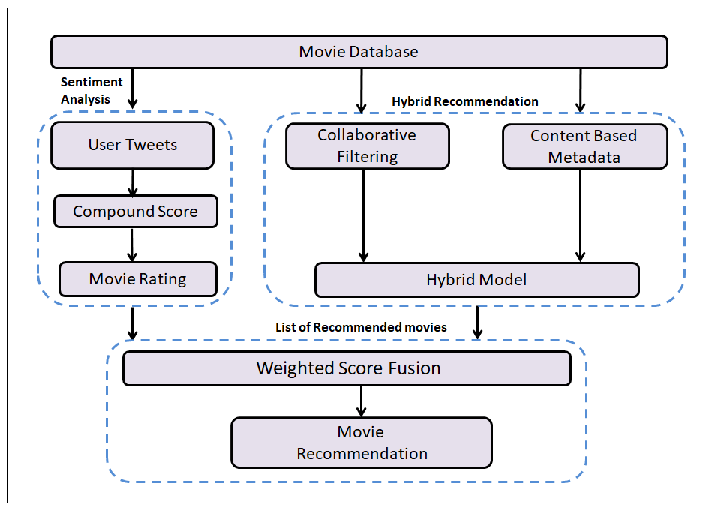}
		\caption{Proposed movie recommendation framework.}
		\label{fig:MABC1}        
	\end{center}
\end{figure}

The proposed sentiment based recommendation system is shown in 
Fig.~\ref{fig:MABC1}. In this section, we describe the different steps and 
components of the proposed recommender system. 
\subsection{Dataset description}
The proposed system needs two different kinds of databases. One is a 
user-rated movies database, where ratings for relevant movies are present and 
another is the user-tweets from Twitter.

\subsubsection{Public database}
There are many popular public databases available which
have been widely used in the area of movie recommendation 
systems. In proposed framework, the tweets were extracted from Twitter 
for movies available in the database. In order to 
incorporate the sentiment analysis in the proposed recommendation system, the tweets 
of relatively new movies were used in this work.


Experiments conducted on public databases such as the Movielens 100K \footnote{https://grouplens.org/datasets/movielens/100k/}, Movielens 20M \footnote{https://grouplens.org/datasets/movielens/20m/}, Internet Movie Database (IMDb\footnote{https://www.kaggle.com/orgesleka/imdbmovies}), and the Netflix database   \footnote{https://www.kaggle.com/netflix-inc/netflix-prize-data/data} were not found suitable for our work. These databases were relatively outdated and contained old movies whose appropriate
microblogging data was not available. After thorough assessment of different databases,
the MovieTweetings database \cite{sdooms} was selected for the proposed system.

MovieTweetings is widely considered as a modern version of MovieLens database. The 
MovieTweetings database is an unfiltered database, unlike MovieLens where a single 
user has rated at least 20 movies. The goal of this database is to provide an 
up-to-date rating so it contains more realistic data for sentiment 
analysis. This database is extracted from social media. It is very diverse but 
has a low sparsity value. Table \ref{my-label1} displays the relevant details of 
the MovieTweetings database.

\begin{table}[h]
\centering
\caption{Details of MovieTweetings database.}
\label{my-label1}
\begin{scriptsize}
\begin{tabular}{l l}
\hline
\textbf{Metric}   & \textbf{Value}                          \\ \hline

\hline
Ratings                               & 646410                              \\ 
Unique Users                          & 51081                               \\ 
Unique Movies                         & 29228                               \\ 
Start Year                            & 1894                                \\
End Year                              & 2017                                \\ 
\hline
\end{tabular}
\end{scriptsize}
\end{table}

\subsubsection{Modified MovieTweetings database}
In our proposed work, the MovieTweetings database is modified for implementing 
the recommendation system. The main motivation for the modification was to use 
sentiment analysis of tweets by users in the prediction of the recommendation 
of movies. The MovieTweetings database contains movies ranging from years 
1894 to 2017. Due to the scarcity of tweets for old movies, we only considered movies that had a release year of more than 2014 and extracted a subset of the MovieTweetings database complying with our motive.
\begin{equation}
 release\_year_{movies} \ge 2014
\end{equation}

After this database processing, a new tweaked database was formed  
for the 
implementation of the recommendation system. This modified database consisted of 
$292863$ ratings by 51081 users on 6209 different movies.



  
 
\begin{figure}[!h!t!b]
	\begin{center}
		\includegraphics[width=.7\linewidth]{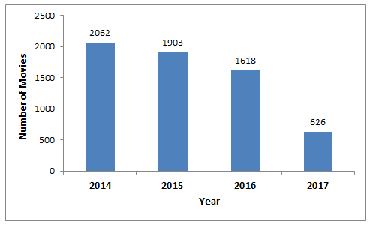}
		\caption{Year wise movies distribution.}
		\label{fig:MABC}        
	\end{center}
\end{figure}

We find that $45$\% of the ratings consist of movies between 2014-2017 which is 
around $20$\% of the total movies of the whole MovieTweetings database So in this work, only movies between 2014 -2017 has been used. The distribution of movies with respect to the year of release is depicted in Fig. \ref{fig:MABC}. 
The MovieTweetings database has three different components. First component 
contains the mapping of users with their Twitter IDs. The second component 
contains the ratings of movie items as rated by the users and the final 
component contains information about the movies that were rated.  

The extracted database 
from MovieTweetings contains the 
ratings of movies by users and their respective genres. 
However, it does not contain any other data except for the release 
year and genres. Such data is 
proficient only in case of social filtering where there are enough users with 
overlapping characteristics in the system. We can use such data in case of 
collaborative filtering where suggestions are made solely based on the ratings 
given by a pertaining user and items are recommended on the judgement of 
similarity between users. In the proposed model, the socially filtered data as well as the similarity of 
movies based on their attributes has been used. The Movie Database (TMDb) API was used to get the attribute of the movies.
TMDb \footnote{https://www.themoviedb.org/about?language=en} is a premier 
source for an extensive metadata for movies that has more than 30 languages. 
The movie attributes of MovieTweetings database are shown in Table~\ref{tab:augmented}.



\begin{table}[h]
\centering
\caption{Example of a movie entry in the modified MovieTweetings database.}
\label{tab:augmented}
\begin{scriptsize}
\begin{tabular}{l l}
\hline
\textbf{Attribute}   & \textbf{Value}                          \\ \hline
MovieID              & 0451279                                 \\ 
Title                & Wonder Woman                            \\ 
Runtime              & 141 min                                 \\ 
Genre                & Action,Adventure,Fantasy                \\
Director             & Patty Jenkins                            \\ 
Writer               & Allan Heinberg                           \\ 
Actors               & Gal Gadot’,Chris Pine                   \\ 
Rating               & 7.6         Massachusetts Institute of Technology in 1996.                             \\ 
Production Companies & DC Films,Tencent Pictures               \\ 
Popularity           & 524.772                                 \\ 
Language             & en                                      \\ 
Production Countries & United States of America                \\ 
Budget               & 816303142                               \\ \hline
\end{tabular}
\end{scriptsize}
\end{table}

The modified database contains very obscure movies from different countries and 
different languages whose metadata was not available 
in TMDb. Such movies were discarded that have almost nil metadata. The final database had around 4500 movies.
Table~\ref{tab:final_dataset} shows the statistical details of the database. 

\begin{table}[h]
\centering
\caption{Details of the modified MovieTweetings database.}
\label{tab:final_dataset}
\begin{scriptsize}
\begin{tabular}{l l}
\hline
\textbf{Metric}   & \textbf{Value}                          \\ \hline
Ratings         & 292863                              \\ 
Unique Users    & 51081                               \\ 
Unique Movies   & 4515                                \\ 
Start Year      & 2014                                \\ 
End Year        & 2017                                \\ \hline

\end{tabular}
\end{scriptsize}
\end{table}

\subsection{Analysis of user tweets}
\label{sec:MPFEC}
\begin{figure}[!h!t!b]
	\begin{center}
		\includegraphics[width=.9\linewidth]{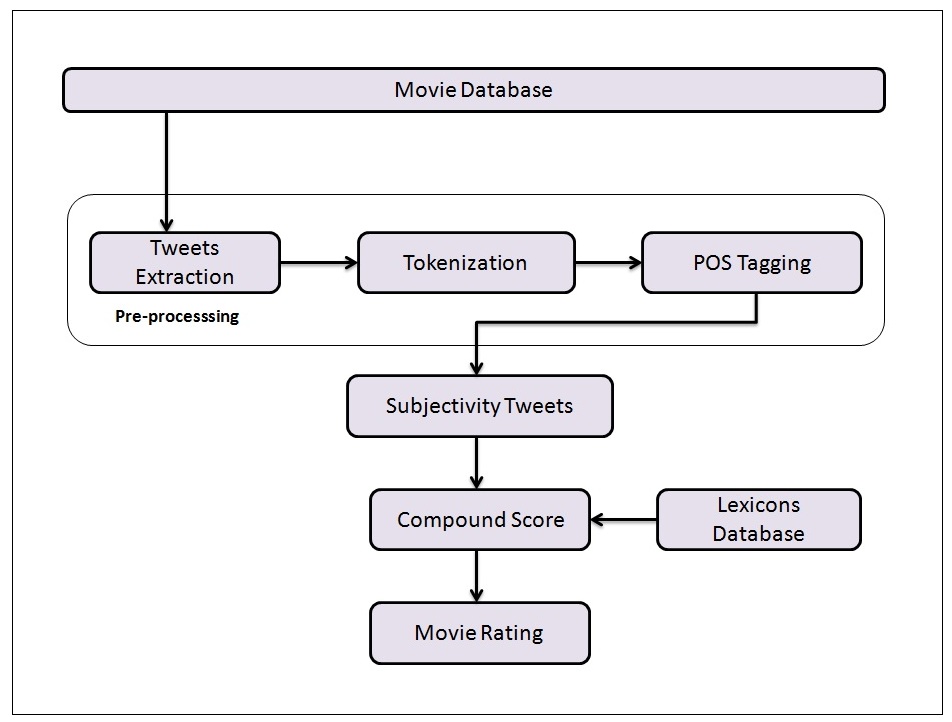}
		\caption{A framework of the VADER sentiment analysis system.}
		\label{fig:MABC2}        
	\end{center}
\end{figure}

The tweets of movies from MovieTweetings were fetched through Twitter API \footnote{https://developer.twitter.com/en/docs} as shown in Fig. \ref{fig:MABC2}.
The extracted tweets consisted tremendous amounts of noise such as hashtag, emoji, repetitive words 
and other irrelevant data which were removed using preprocessing techniques.

%
%

\subsubsection{Preprocessing of tweets}
 
There were many noisy and uninformative parts in tweets, which did not contribute much to accurate 
sentiment analysis. These include stop words, punctuations, web links, special characters and repetitive words. Few example of noisy and uninformative parts are mentioned in Table \ref{tab:ENRD}. After preprocessing, the text extracted from the tweets was used for sentiment analysis and then this information was used for  
building the recommendation system. 

\begin{table}[!h!t!b]
\centering
\caption{Examples of noisy and uninformative parts in tweets.}
\label{tab:ENRD}
\begin{scriptsize}
\begin{tabular}{l l}
\hline
\textbf{Types  of noise}   & \textbf{Example}                          \\ \hline
	
Stop words                    & a, and, the, after, am    \\ 
Lemma                         & serve, served and serving \\ 
Web links                     & www.tripadvisior.com      \\ 
Filtering of repeating words & happyyyy, heloooo          \\ 
Special Characters             & !, @, \#, \$, \%, and \_  \\ \hline
\end{tabular}
\end{scriptsize}
\end{table}

\subsubsection{Sentiment analysis of user tweets}

VADER is a computationally efficient lexicon of sentiment-related words and rule-based method that is used for sentiment analysis. In  this  approach,  the  
detection of subjectivity tweets can  be  handled  by  the  sentiment process itself. VADER maps words to sentiment by building a lexicon or a 
‘dictionary of sentiments’. The dictionary evaluates the sentiment of phrases and 
tweets. Lexical approaches can be employed to evaluate either the sentiment category or the score of the tweets.
From each tweet, VADER produces four sentiment components. Out of these the first three are: positive, 
neutral, and negative. The fourth component is a compound score which is a 
normalized score of first three metrics. The range of compound score varies from 
-1 to 1, where -1 represents least preferred and 1 denotes most preferred choice 
of movies. Later, the score is scaled in the range $1-10$ using Eq. (\ref{eq:fit1}).  

\begin{equation}
\label{eq:fit1}
Rating=[1+\left(1+x\right)\times2]\times2    
\end{equation}

where $x$ is a compound score. 

In addition, VADER executes faster in comparison to other methods such as 
Naive Bayes analyzer, TextBlob. They work well in texts having slangs, emoticons, and 
acronyms. 


\subsection{Hybrid recommendation}
\label{sec:meth1}
In this section, we describe the combination of content-based similarity features with 
collaborative social filtering to generate a hybrid recommendation model. Let 
$f$ = \{$f_1$,$f_2$,$\cdots$,$f_n$\} and $q$ = \{$q_1$,$q_2$,$\cdots$,$q_n$\} are the content-based feature vectors and  weight vector, respectively.

We construct the closeness \textit{C} of two items \textit{$i$} and 
\textit{$j$} as:

\begin{equation}
\label{eq:closeness}
C(i,j) = 
  \begin{cases} \sum_{n=1}^{N} f_n(A_{n_{i}},A_{n_{j}}), & \text{for\ 
} i\neq j \\
	0, & otherwise
  \end{cases}
\end{equation}

where $f_n(A_{n_{i}},A_{n_{j}})$, corresponds to the similarity between feature 
values $A_{n_{i}}$ and $A_{n_{j}}$ corresponding to two items. In Eq. (\ref{eq:closeness}) the closeness of the items is determined using the metadata or the 
relevant information related to the items. $F_{ij}$ are constructed  by 
combining the closeness vector $C$ for all the items and multiplying it with the weight vectors $q$. $F_{ij}$ is a feature 
matrix of dimension $n \times \frac{M(M-1)}{2}$, where $n$ and $M$ are the number of 
feature attributes and number of items, respectively.   


The weight vectors $q$ are evaluated using a social graph of items that indicate the user likeness of items. Let $U = \{u_1,u_2,\cdots,u_n\}$ 
where $u_i$ is a user in the database. A User-Item matrix is constructed for $M$ items. An important property of the User-Item 
matrix is that it has very high sparsity. Typical collaborative Filtering ~\cite{collab} 
uses this User-Item matrix to predict a user's rating of a particular item $i$ 
by analyzing the ratings of other users in the user's neighborhood, normally a 
K neighboring users. Neighboring users are recognized by similarity measures 
like Cosine Similarity, Pearson Correlation etc. After selecting K neighbouring users, the 
weighted aggregation of the ratings are as follow:

\begin{equation}
\label{eq:final_eq1}
 rating(u,i) = \frac{1}{K} \sum_{k \in K} similarity(user_u,user_{v_{k}}) \cdot 
rating_{v_{ki}}
\end{equation}

where $u$ and $v_k$ are target user and K nearest neighbors, respectively.
The procedure of collaborative filtering is used to overcome the sparsity of the 
User-Item matrix instead of directly using it to predict ratings. We employ the 
tweaked User-Item matrix to construct a social graph using items as nodes. This 
graph represents user perception of similarity between items. The determination 
of feature weights complies with the social graph. 

In order to determine the optimal feature weights $q$, we formulate a framework as described in  
Eq. (\ref{eq:one_closeness}):
\begin{equation}
\label{eq:one_closeness}
 S(i,j) = q \cdot F_{ij}
\end{equation}

which can be expanded as:
\begin{equation}
\label{eq:final_eq}
S(i,j)=q_1\cdot f_1(A_{1i},A_{1j})+q_2\cdot 
f_2(A_{2i},A_{2j})+\cdots+q_n \cdot f_n(A_{ni},A_{nj})
\end{equation}

Now for two items $i$ and $j$, $S(i,j)$ 
is evaluated from the User-Item matrix, where $S(i,j)$ are the number of 
users who are interested in both items $i$, $j$ . For the entire database, $S$ is 
a matrix of dimension $1 \times \frac {M(M-1)}{2}$ and $q$ is a matrix of dimension $1 
\times n$, where $n$ is the number of content-based features and 
dimenstionality of $F$ is $n \times \frac {M(M-1)}{2}$. We calculate the weight 
vector $q$ for all the metadata feature attributes for the complete items using 
the Moore-Penrose Pseudoinverse as in Eq. (\ref{eq:feat_weights}) :

\begin{equation}
\label{eq:feat_weights}
 q = S^{-1} \cdot F
\end{equation}

\subsection{Weighted score fusion}
 
In section \ref{sec:meth1}, we derived the weights $q$ of the feature vectors using 
both movie-similarity and user-similarity paradigms. These weights $q$ are normalized between [0, 1] and the concept of sentiment-fusion is utilized in the proposed system. 
Through the retrieved user tweets, a sentiment rating is fabricated for all movies $M$. 
Let $S \in \{s_1,s_2,\cdots,s_n\}$ where $s_i$ is the rating  
of movie i calculated using Eq. (\ref{eq:fit1}). A function $G(i,j)$ 
for two movies $i$, $j$ are defined based on their sentiment ratings $s_i$ and $s_j$ as mentioned in Eq. (\ref{eq:sent_score}):

\begin{equation}
 \label{eq:sent_score}
 G(i,j) = D - | s_i - s_j | 
\end{equation}
where D is a constant. The constant D in Eq. (\ref{eq:sent_score}) is taken as $10$ because the ratings are in a scale of $1-10$. 
Another function $H(i,j)$ defined as:

\begin{equation}
 \label{eq:items_simil}
 H(i,j) = q\cdot f_{ij}
\end{equation}

where $f_{ij}$ is the feature similarity between movies $i, j$ and $q$ are the set of optimal weights as determined by Eq. (\ref{eq:feat_weights}).
The final combined similarity $CS(i,j)$ is described in Eq. (\ref{eq:final}). It is a 
weighted combination of the defined functions $G$ and $H$. 

\begin{equation}
\label{eq:final}
 CS(i,j) = \omega_1 \cdot H(i,j) + \omega_2 \cdot G(i,j)
\end{equation}
\begin{equation}
\label{eq:weights}
 \omega_1 + \omega_2 = 1, \qquad \qquad \omega_1, \omega_2 \in [0,1]   
\end{equation}

where $\omega_1$ corresponds to the weight of the similarity score calculated from the 
hybrid model and $\omega_2$ corresponds to the weight of the sentiment similarity score.

\section{Experimental results and analysis}
\label{sec:results}

In this section, the different correlation coefficient results are presented between 
sentiment ratings and IMDb movie ratings. We also describe a comprehensive quantitative and qualitative analysis, illustrating the efficacy and the precision of our proposed framework on movie database. 

\subsection{Correlation between sentiment and IMDb movie ratings}

We conducted the statistical analysis between sentiment ratings $X$ and movie rating $Y$ to find the correlation coefficient. The correlation coefficient value varies from -1 to +1. Let $D$ denote a database of movies and $N$ denotes the number of total movies in the database. The statistical correlation coefficients are as follows: Spearman Rank Order, Correlation Coefficient ($SROCC$), Kendall Rank Correlation Coefficient ($KRCC$) and Pearson Linear Correlation Coefficient ($PLCC$). Table \ref{corr} displays the values of different co-relation coefficients utilized by us. In our experiments, we  have found that sentiment and movie ratings are positively correlated. For PLCC, $x_i$ and $y_i$ are sentiment rating and IMDb movie rating, respectively for $i^{th}$ movie while $\bar{x}$ denotes the mean sentiment score and $\bar{y}$ denotes the mean movie rating in the database. For SROCC, $d_i$ is the difference between the sentiment rating and movie rating of the $i^{th}$ movie in the database. For KRCC, $N_c$ and $N_d$ represents the number of concordant and discordant pairs in the database, respectively.

\begin{table}[!htp]
\centering
\caption{Correlation measures between sentiment and movie ratings.}
\label{corr}
\begin{tabular}{c c c}
\hline
\textbf{Correlation coefficient} & \textbf{Definition} & \textbf{Value} \\ \hline                                                                                                                                                                                                                                                                                                                                                                                                                                                                                                                                      
PLCC                 & $ \frac{\sum_{i=1}^N(x_i-\bar{x})(y_i-\bar{y})}{\sqrt{\sum_{i=1}^N(x_i-\bar{x})^2}\sqrt{\sum_{i=1}^N(y_i-\bar{y})^2}} $  & 0.76  \\ 
SROCC                & $ 1-\frac{6}{N(N^2-1)}\sum\limits_{i=1}^Nd_i^2 $  & 0.72 \\                                                                                                                                                                                                                                                                                                                                                                                                                                                                                                                                             
KRCC                 & $ \frac{2(N_c-N_d)}{N(N-1)} $  & 0.51  \\ \hline                                                                                                                                                                                                                                                                                                                                                                                                                                                                                                                                           
\end{tabular}
\end{table}


%
%

\subsection{Evaluation metric}
\label{subsec:metric}

In many real-world applications, relevant recommendations are suggested by the system, instead of directly predicting rating values. This is known as \textit{Top-N} recommendations ~\cite{topn,eval,said2013top} and suggests specific items to users that are likeable. Normally, in majority of the common literature RMSE (Root Mean Squared Error) or MSE (Mean Squared Error) are used. But such error metrics are less significant for \textit{Top-N} recommendation system. Therefore, the direct alternative methodologies are used for evaluation metric (e.g. precision). Precision is defined in terms of movies that are relevant and the movies that are recommended by the model. Let $L_{rel}$ be the relevant movies and $L_{rec}$ be the recommended movies. In proposed system, Precision@N is defined as follow in Eq. (\ref{eq:prec}):

\begin{equation}
 \label{eq:prec}
 Precision@N = \frac{L_{rel} \cap L_{rec}}{L_{rec}}
\end{equation}


For the proposed model, the choice of weights in the fusion in Eq. 
(\ref{eq:final}) is determined by evaluating the Precision@5 and Precision@10 for 
a different combination of weights $\omega_1$ and $\omega_2$ conforming with Eq. 
(\ref{eq:weights}).

\subsection{Weight selection for weighted fusion}
\label{subsec:metric1}
For every movie, \textit{Top-N} recommendation list is evaluated using Eq. (\ref{eq:final}). The choice of the weights $\omega_1$ and $\omega_2$ in Eq. (\ref{eq:final}) are decided by experiments conducted on the metric mentioned in the section ~\ref{subsec:metric}. 
The Precision@N is evaluated as in Eq. (\ref{eq:prec}). The recommendations of all movies are collected from public databases like IMDb and TMDb. The recommended movies by these two sources are consider as the ground-truth. We compare the results of the Precision@5 and Precision@10 for different values of $\omega_1$ and $\omega_2$. We choose the values of $\omega_1$ and $\omega_2$ for which the precision values are the best. 

\begin{figure}[!h!t!b]
	\begin{center}
		\includegraphics[width=0.8\linewidth]{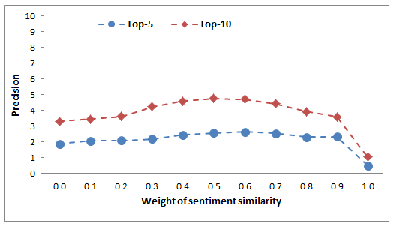}
		\caption{Precision of Top-5 and Top-10 movies with varying sentiment 
similarity weights.}
		\label{fig:precision}        
	\end{center}
\end{figure}


From Fig. \ref{fig:precision}, an observation can be made that 
the maximum precision for weight values are between $0.5$ and $0.6$. Hence, $\omega_1$ and $\omega_2$ values are selected 0.5 in the proposed system.

\subsection{Comparative analysis}

\begin{figure}[!h!t!b]
	\begin{center}
		\includegraphics[width=.9\linewidth]{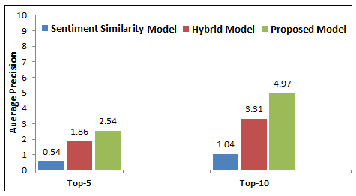}
		\caption{Comparison of proposed model with baselines models.}
		\label{fig:comp}        
	\end{center}
\end{figure}

In this section, we present a comparative analysis of our proposed system with base-line models. The base-line models used for the purpose of our experiments are  Pure Hybrid Model and  Sentiment Similarity model. The  Pure Hybrid Model (PH Model) is a combination of content-based filtering and collaborative filtering where both these filtering methods supplement each other to increase the robustness of the complete model. The recommended movies are based on the similarity of attributes like the genre,  director, actor, etc. The similarities are evaluated using weights obtained using a social graph as described in Section \ref{sec:meth1}.  Sentiment Similarity model (SS Model) recommends the movie based solely on the similarity of the movie tweets of the corresponding tuple of movies. We evaluate our proposed method using Precision@5 and Precision@10. Fig.~\ref{fig:comp} displays the quantitative comparative results of our proposed system with the baseline models. For Precision@5, the average precision value of SS Model and PH Model are 0.54  and  1.86,  respectively. Similarly, for Precision@10,  the average precision value of SS Models and PH Model are 1.04 and 3.31, respectively. Our proposed model achieves a  better precision value in both cases with 2.54 for \textit{Top-5} and 4.97  for \textit{Top-10} in comparison to  PH  and  SS  Model. Thus we can infer that our method will suggest at least 2 recommended movies out of 5 and 5 recommended movies out of 10 considering the general consensus as validated by IMDb and TMDb databases.

\subsection{Qualitative analysis}
\label{sec:Qualitative Results}

In this section, we show the qualitative results of some of the recommended movies given by the proposed system. It also includes movies from other languages as shown in the Tables \ref{wonderwoman} and \ref{neerja}. It is clear from these tables that the 
recommendations from the proposed system have many intersecting movies 
both with the recommendations from IMDb and TMDb. The qualitative analysis of  Hollywood as well as Bollywood (Indian) movies show the capability of suggesting recommendations in diverse situations. 
 

\begin{table}[]
\centering
\caption{Qualitative analysis of \textit{Wonder Woman} movie: Language (English). Movies in \textbf{bold} are intersecting 
with either IMDb or TMDb}
\label{wonderwoman}
\begin{scriptsize}
\begin{tabular}{l l l}
\hline
\textbf{IMDb}                      & \textbf{TMDb}                 
                                                & 
\textbf{\begin{tabular}[c]{@{}l@{}}Recommendations from the \\ proposed 
system\end{tabular}} \\ \hline
\textbf{Justice League}                     & \textbf{Guardians of the Galaxy Vol. 2}              
                                                & \textbf{Batman v Superman: Dawn of 
Justice}\\ 
\textbf{Batman v Superman: Dawn of Justice} & Spider-Man:  Homecoming                     
                                                & \textbf{Suicide Squad}                  
                                                              \\ 
\textbf{Suicide Squad                      }& Logan                                       
                                                & \textbf{Thor: Ragnarok}
                                                              \\ 
\textbf{Thor: Ragnarok                     }& \textbf{Thor: Ragnarok                              }
                                                & \textbf{Justice League}        
                                                              \\ 
Spider-Man: Homecoming             & \textbf{Justice League                              }
                                                & Warcraft                       
                                                              \\ 
Deadpool                           & \begin{tabular}[c]{@{}l@{}}Pirates of the 
Caribbean: \\ Dead Men Tell No Tales\end{tabular} & \textbf{Doctor Strange}               
                                                                \\ 
Logan                              & \textbf{Doctor Strange                              }
                                                & \textbf{Guardians of the Galaxy Vol. 2}
                                                              \\ 
Captain America: Civil War         & Baby Driver                                 
                                                & \textbf{Kong: Skull Island             }
                                                              \\ 
\textbf{Doctor Strange                     }& \textbf{Kong: Skull Island                          }
                                                & The LEGO Batman Movie                         
                                                              \\ 
\textbf{Guardians of the  Galaxy Vol. 2    }& Life                                        
                                                & Batman and Harley Quinn        
                                                              \\ \hline
\end{tabular}
\end{scriptsize}
\end{table}

\begin{table}[!h!t!b]
\centering
\caption{Qualitative analysis of \textit{Neerja} movie: Language (Hindi). Movies in \textbf{bold} are intersecting 
with either IMDb or TMDb}
\label{neerja}
\begin{scriptsize}
\begin{tabular}{l l l}
\hline
\textbf{IMDb}                & \textbf{TMDb} & 
\textbf{Recommendations from the proposed system} \\ \hline
\textbf{Airlift                      }& \textbf{Airlift                     }& \textbf{Simran}
                              \\ 
Pink                         & Pink                        & \textbf{Fan}
                              \\ 
Kapoor \& Sons               & Rustom                      & \textbf{Raabta}              
                              \\ 
\textbf{Udta Punjab                  }& Ghayal Once Again           & \textbf{Udta Punjab}         
                              \\ 
Drishyam                     & Mary Kom                    & \textbf{Rocky Handsome}
                              \\ 
Rustom                       & \textbf{Udta Punjab                 }& \textbf{Rangoon}             
                              \\ 
M.S. Dhoni: The Untold Story & \textbf{Force 2                     }& \textbf{Raabta}
                              \\ 
\textbf{Raabta                       }& \textbf{Fan                         }& \textbf{Force 2}             
                              \\ 
Dear Zindagi                 & \textbf{Rocky Handsome              }& Te3n                
                              \\ 
\textbf{Rangoon                      }& \textbf{Simran                      }& \textbf{Airlift}
                              \\ \hline
\end{tabular}
\end{scriptsize}
\end{table}

\section{Conclusion and future works}
\label{sec:conc}
Recommender systems are an important medium of information filtering system in the modern age where there are enormous amounts of data readily available. In this paper, we have proposed a recommender system which uses sentiment analysis data from twitter along with movie metadata and a social graph to recommend movies. Sentiment analysis provide information about how audience is reacting to a particular movie and this information has been observed to be useful. The proposed system used weighted score fusion to improve the recommendations. Based on our experiments, the average precision in \textit{Top-5} and \textit{Top-10} for sentiment similarity, hybrid, and proposed model are .54 \& 1.04, 1.86 \& 3.31, and 2.54 \& 4.97, respectively. We found that proposed model recommends more precisely than the others models. In future, we look forward to extract more information about emotional tone of the user from different social media platforms to further improve the recommendation system using sentiment analysis. The findings of this paper show that 
there is enough scope for research to explore emotion information about users and integrate them in recommendation system for e-commerce application. Another direction of research will be to update weights derived from the social graph in real-time. We have used a static database for our experiments where we only considered movies up to 2017. This framework can be explored in a dynamic paradigm where the inclusion of recent movies could be made regular.

\section*{Compliance with ethical standards}
\textbf{Conflict of interest} The authors declared that they have no conflicts of interest to this work.

\bibliographystyle{plainnat}       

\bibliography{egbib}
%
%

\end{document}